\documentclass[a4paper]{jpconf}
\usepackage{graphicx,color}

\newcommand{\dvol}{\mbox{dvol}}

\begin{document}

\title{Spherical steady-state accretion of a relativistic collisionless gas into a Schwarzschild black hole}

\author{Paola Rioseco and Olivier Sarbach}

\address{Instituto de F\'isica y Matem\'aticas,
Universidad Michoacana de San Nicol\'as de Hidalgo,
Edificio C-3, Ciudad Universitaria, 58040 Morelia, Michoac\'an, M\'exico.}

\ead{prioseco@ifm.umich.mx, sarbach@ifm.umich.mx}

\begin{abstract}
In previous work, we derived the most general solution of the collisionless Boltzmann equation 
describing the accretion of a kinetic gas into a Schwarzschild black hole background, and we gave explicit expressions for the corresponding observables (the current density and stress energy-momentum tensor) in terms of certain integrals over the distribution function. In this article, we numerically compute these integrals for the particular case of the steady-state, spherical symmetric accretion flows which, at infinity, are described by an equilibrium distribution function of given temperature. We analyze in detail the behavior of the observables as a function of the temperature and the radial coordinate, comparing our results with the perfect fluid model of Bondi-Michel accretion.
\end{abstract}

\section{Introduction}

During the last few years there has been an increasing interest in the relativistic kinetic theory of gases, originating from the need to consider kinetic gases in extreme situations, either at very high temperatures or in the presence of strong gravitational fields, where the standard Newtonian theory breaks down. Such situations typically occur in astrophysical or cosmological scenarios, for instance in the vicinity of black holes or in the early Universe. For a recent book on relativistic kinetic theory, we refer the reader to~\cite{CercignaniKremer-Book}. Other relevant reviews include the book by Stewart~\cite{Stewart-Book}, the survey article by Ehlers~\cite{jE71}, and the survey article by Andr\'easson~\cite{hA11} on more recent mathematical applications of relativistic kinetic theory. Incorporating the principle of general covariance of Einstein's theory of general relativity, the general relativistic kinetic theory of gases possesses an elegant geometric formulation based on the tangent (or cotangent) bundle associated with the spacetime manifold $(M,g)$, see for example Refs.~\cite{rL66,jE71,jE73,oStZ13,oStZ14a,oStZ14b,pRoS16}.

In previous work~\cite{pRoS16}, we provided a systematic study for the propagation of a collisionless, relativistic kinetic simple gas on a nonrotating black hole background, neglecting its self-gravity. To this end, we used tools from the theory of integrable Hamiltonian systems and derived the most general solution of the collisionless Boltzmann equation on a Schwarzschild background describing accretion. Further, we derived explicit expressions for the observables (namely, the current density and the stress energy-momentum tensor) and specialized them to the particular case of a spherical, steady-state flow. Assuming that asymptotically, the gas is isotropic and in thermodynamic equilibrium at some temperature $T > 0$, we computed the accretion and compression rates, the energy density and the radial and tangential pressures at the horizon in the limit in which the thermal energy $k_B T \ll m c^2$ is much smaller than the rest energy $m c^2$ of the gas particles.\footnote{As usual, $k_B$ denotes Boltzmann's constant and $c$ the speed of light.} In this limit, we found that the tangential pressure at the horizon is about ten times larger than the radial one. This provides a partial explanation for the known fact that the accretion rate for a collisionless gas is low compared to the Bondi-Michel accretion model in the hydrodynamic, isotropic perfect fluid case~\cite{hB52,fM72,Shapiro-Book,eCoS15a}. Furthermore, in~\cite{pRoS16} we showed an asymptotic stability result, proving that rather general initial gas configurations (which are not necessarily stationary nor spherically symmetric) relax in time to a spherical steady-state solution.

Our intention here is to analyze the behavior and properties of the observables corresponding to the spherical steady-state models in more detail. Whereas in our previous work we only analyzed the low temperature limit and the behavior of the observables at the horizon and in the asymptotic region, here we extend the analysis to arbitrary temperatures and values of the radial coordinate. This is achieved by computing numerically the integrals over the momenta derived in~\cite{pRoS16} and plotting the results. By interpreting these plots we obtain a better understanding for the physical properties of a collisionless kinetic gas which is accreted by a black hole and the manner it differs from the behavior of a perfect fluid in the same situation. In the last part of the article, we provide an explanation for the reason why the kinetic gas ceases to behave as an isotropic perfect fluid in the vicinity of the black hole, even though in our model it fulfills the same conditions as a perfect fluid at infinity.

\section{The spherical steady-state model}

Our (highly idealized) model of black hole accretion is based on the assumptions that the accretion flow is steady-state, spherically symmetric, and that the self-gravity and collisions of the kinetic gas being accreted can be neglected. In the cotangent-bundle formulation the gas configuration can be described by a one-particle distribution function $f(x,p)$ which satisfies the collisionless Boltzmann equation\footnote{Here, $(x^\mu,p_\mu)$ refer to adapted local coordinates on the cotangent bundle $T^* M$, in which the canonical momentum is expanded as $p = p_\mu dx^\mu$.} 
\begin{equation}\label{Liouville}
\left( p^\mu\frac{\partial}{\partial x^\mu} - \frac{1}{2} p_\alpha p_\beta\frac{\partial g^{\alpha\beta}}{\partial x^\mu} \frac{\partial}{\partial p_\mu} \right) f = 0
\label{Eq:LiouvilleEq}
\end{equation}
on a Schwarzschild black hole background $(M,g)$. The spacetime observables, that is, the current density and stress energy-momentum tensor, are obtained from the distribution function by integration over the momentum space $C_x$ consisting of those $p$ which have the form $p = g_{\mu\nu}(x) p^\nu dx^\mu$ with $p^\nu\partial_\nu$ denoting a future-directed timelike tangent vector at $x$:
\begin{eqnarray}
J_\mu(x) &=& \int\limits_{C_x} p_\mu f(x,p) \dvol_x(p),\\
T_{\mu\nu}(x) &=& \int\limits_{C_x} p_\mu p_\nu f(x,p) \dvol_x(p),
\end{eqnarray}
with $\dvol_x(p) := \sqrt{-\det(g^{\mu\nu}(x))} d^4 p$ the natural volume element on $C_x$.

In~\cite{pRoS16} we derived the most general solution of Eq.~(\ref{Eq:LiouvilleEq}) in terms of appropriate symplectic coordinates on the cotangent bundle $T^* M$ associated with the Schwarzschild spacetime $(M,g)$ and provided a detailed discussion of the symmetries of $f$. This was achieved by exploiting the fact that geodesic motion in a Schwarzschild spacetime constitutes an integrable Hamiltonian system. By imposing spherical symmetry and stationarity, we showed that any steady-state, spherical configuration is described by a distribution function of the form
\begin{equation}
f(x,p) = {\cal F}(m,E,L),
\label{Eq:SphSol}
\end{equation}
with ${\cal F}$ a smooth function of the mass $m$, energy $E$, and total angular momentum $L$ of the gas particles, which, in terms of standard Schwarzschild coordinates $(t,r,\vartheta,\varphi)$ are given by
\begin{equation}
m = \sqrt{-p^\mu p_\mu},\qquad
E = -p_t,\qquad
L = \sqrt{p_\vartheta^2 + \frac{p_\varphi^2}{\sin^2\vartheta}}.
\end{equation}
Since these quantities are conserved along geodesics, it is clear that Eq.~(\ref{Eq:SphSol}) provides a solution of the collisionless Boltzmann equation. The arguments presented in~\cite{pRoS16} show that any steady-state, spherical solution must be of this form.

For the following, we consider the situation in which there is a reservoir of identical particles at infinity. We assume that these particles have positive rest mass $m > 0$ and that their momenta are isotropically distributed. Additionally, we assume that (although the gas is assumed to be collisionless) some physical process in the past drove the reservoir to thermodynamic equilibrium at some fixed temperature $T > 0$. As a consequence of~(\ref{Eq:SphSol}), this implies that the distribution function is given by
\begin{equation}
f(x,p) = \alpha\delta( \sqrt{-p^\mu p_\mu} - m) e^{-z\varepsilon},
\label{Eq:SphEquilSol}
\end{equation}
where $\alpha$ is a  positive amplitude with untis $1/(\mbox{volume}\times\mbox{mass}^4\times\mbox{velocity}^4)$, $\varepsilon := E/(mc^2)$ is the ratio between the total energy of the particle and its rest energy, and $z$ is the dimensionless inverse temperature
\begin{equation}
z := \frac{mc^2}{k_B T}.
\end{equation}
As stated in the introduction, in~\cite{pRoS16} we explicitly computed the associated observables at the horizon and in the asymptotic region in the low temperature limit $z\to \infty$, and showed that the accretion rate agrees with known results based on Newtonian calculations~\cite{ZelNovik-Book,Shapiro-Book}. In the following, we extend these results to arbitrary values of $z$ and the dimensionless radial coordinate $\xi := 2r/r_H$, with $r_H$ the event horizon radius.

Even though this accretion model is highly idealized and based on a number of assumptions that eventually need to be relaxed, it exhibits rich phenomenology as we show next.

\section{Phenomenology of the spherical steady-state model}

Explicit expressions for the particle and energy fluxes, the current density, and the stress energy-momentum tensor for a distribution function of the form~(\ref{Eq:SphEquilSol}) in terms of integrals over the variable $\varepsilon$ are given in Section IV.C of Ref.~\cite{pRoS16}. In the following, we compute these integrals numerically and analyze their dependency on $z$ and $\xi$.

The particle accretion rate (number of particles that cross the horizon per unit time) and the energy accretion rate (total energy crossing the horizon per unit time) are given by
\begin{eqnarray}
\dot{\cal N} &=& \alpha m^4 c^5\pi^2 r_H^2\int\limits_1^\infty \lambda_c(\varepsilon)^2 e^{-z\varepsilon} d\varepsilon,\\
\dot{\cal E} &=& \alpha m^5 c^7\pi^2 r_H^2\int\limits_1^\infty \varepsilon\lambda_c(\varepsilon)^2 e^{-z\varepsilon} d\varepsilon,
\end{eqnarray}
respectively. In these expressions,
\begin{equation}
\lambda_c(\varepsilon) = \left. 
\sqrt{ \frac{12}{1 - 4a - 8a^2 + 8a\sqrt{a^2 + a}} } \right|_{a = \frac{9}{8}\varepsilon^2 - 1},
\label{Eq:lambdac}
\end{equation}
is the critical value of the (dimensionless) total angular momentum, below which the particles are absorbed by the black hole. For a derivation of Eq.~(\ref{Eq:lambdac}) and details, see Appendix A in~\cite{pRoS16}. For the purpose of this article it is enough to know that $\lambda_c(\varepsilon)$ is a monotonously increasing function such that $\lambda_c(1) = 4$ and $\lambda_c(\varepsilon) \simeq \sqrt{27}\varepsilon$ for $\varepsilon\to \infty$.

\begin{figure}[ht]
\centerline{\resizebox{8.0cm}{!}{\includegraphics{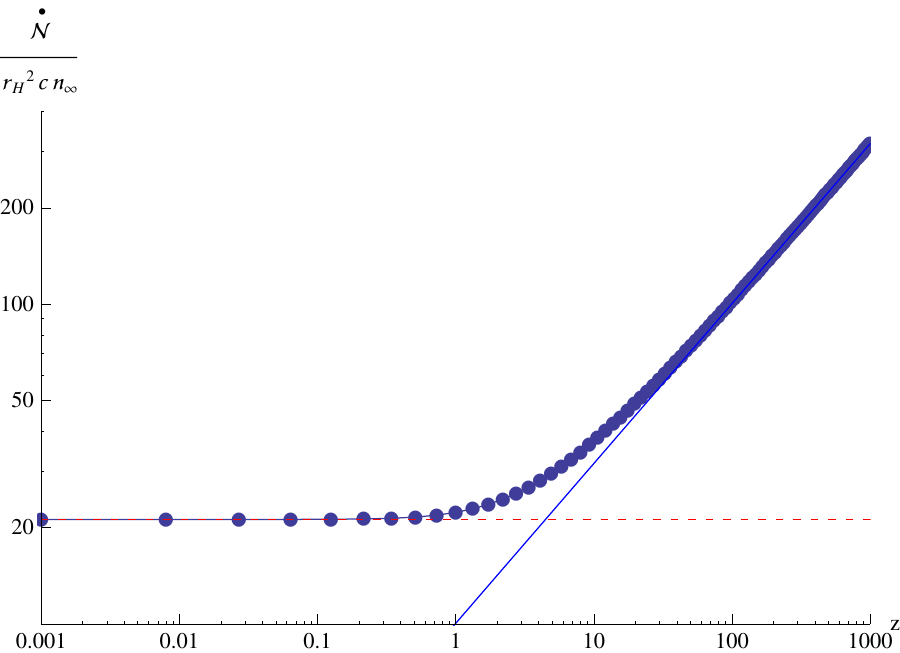}}\resizebox{8.0cm}{!}{\includegraphics{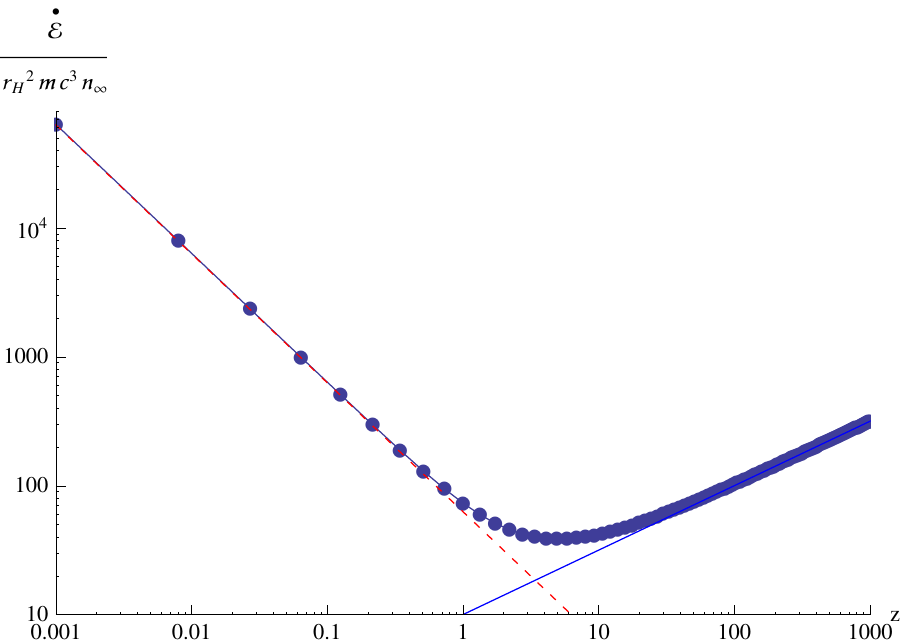}}}
\caption{\label{Fig:AccretionRate} Particle and energy accretion rates as a function of the inverse temperature $z$. The solid blue lines describe the asymptotic behavior in the low temperature limit $z\to \infty$, see Eqs.~(\ref{Eq:NdotLowT},\ref{Eq:EdotLowT}). The red dashed lines describe the behavior in the high temperature limit, see Eqs.~(\ref{Eq:NdotHighT},\ref{Eq:EdotHighT}).
}
\end{figure}

For very low temperatures ($z\to \infty$), $\dot{\cal E} \simeq mc^2\dot{\cal N}$, since each particle's energy $E \simeq m c^2$ is dominated by its rest mass energy. However, for finite values of $z$ the energy accretion rate $\dot{\cal E}$ is larger than $m c^2\dot{\cal N}$ since it includes the internal energy of the gas. When analyzing the behavior of these accretion rates as a function of temperature, it is convenient to eliminate the unphysical parameter $\alpha$. This can be done by replacing $\alpha$ with the particle density at infinity, given by~\cite{fJ11a,fJ11b,wI63} 
\begin{equation}
n_{\infty}(z) = 4\pi \alpha (m c)^4\frac{K_2(z)}{z},
\label{Eq:ninfinity}
\end{equation}
where $K_2$ is the modified Bessel function of the second kind. In Fig.~\ref{Fig:AccretionRate} we plot the dimensionless quantities $\dot{\cal N}/(r_H^2 c n_\infty)$ and $\dot{\cal E}/(r_H^2 mc^3 n_\infty)$ as a function of the dimensionless inverse temperature $z$.

For low temperatures one finds~\cite{pRoS16}
\begin{eqnarray}
\dot{\cal N} &\sim& 4r_H^2 c \sqrt{2\pi z} n_\infty \simeq
 (8.3 \times 10 ^{28}) \left(  \frac{M_{H}}{10M _{\odot}} \right)^2\left( \frac{n_\infty}{1cm^3} \right)\left( \frac{z}{10^9} \right)^{1/2} \frac{1}{s}, 
\label{Eq:NdotLowT}\\
\dot{\cal E} &\sim& 4r_H^2  mc^3 \sqrt{2\pi z} n_\infty 
\simeq (2.2 \times 10^{-21}) \left(  \frac{M_{H}}{10M _{\odot}} \right)^2\left(\frac{m}{m_p}\right)\left( \frac{n_\infty}{1cm^3} \right)\left( \frac{z}{10^9} \right)^{1/2} \frac{M_\odot c^2}{yr},
\label{Eq:EdotLowT}
\end{eqnarray}
with $M_\odot$ and $m_p$ denoting the solar and proton mass, respectively, and where we have used typical values for the ionized component of the interstellar medium in our Galaxy. This result agrees with Newtonian-based calculations~\cite{ZelNovik-Book,Shapiro-Book}, and it differs from the corresponding result in the isotropic perfect fluid case by a factor of the order of $z$, see for instance~\cite{Shapiro-Book}. In the opposite limit of extremely high temperatures we find, using a similar method than the one described in Appendix C of~\cite{pRoS16},
\begin{eqnarray}
\dot{\cal N} &\simeq& \frac{27\pi}{4} r_H^2 c n_\infty,
\label{Eq:NdotHighT}\\
\dot{\cal E} &\simeq& \frac{81\pi}{4} \frac{r_H^2 m c^2 n_\infty}{z}.
\label{Eq:EdotHighT}
\end{eqnarray}
We see that while the particle accretion rate converges to a constant for small $z$, the energy accretion rate diverges like $1/z$.

At this point, the question arises why the particle accretion rate $\dot{\cal N}$ decreases steadily as the temperature increases, while the energy accretion rate $\dot{\cal E}$ starts decreasing until it reaches a minimum, located about $z\sim 1$, and then increases again. This difference can be understood as follows. For small temperatures most of the particles in the reservoir have very low internal energy, and thus most of the infalling particles are absorbed by the black hole, leading to a high particle accretion rate (compared to $r_H^2 c n_\infty$). In contrast, at high temperatures, only a small fraction of the infalling particles will be absorbed by the hole, since most of the gas particles have large enough angular momenta $\lambda > \lambda_c(\varepsilon)$. To make this point more precise, we consider the critical impact parameter below which the particle is absorbed by the black hole, which is found to be
\begin{equation}
b_c(\varepsilon) = \frac{L_c(E)}{\sqrt{E^2-m^2}} 
 = \frac{r_H}{2}\frac{\lambda_c(\varepsilon)}{\sqrt{\varepsilon^2-1}}.
\end{equation}
For small temperatures, most particles have energy $\varepsilon\simeq 1$ yielding large values of $b_c(\varepsilon)$, showing that in this case most particles are absorbed by the black hole. In contrast, at high temperatures most particles have energies $\varepsilon\gg1 $ in which case $b_c(\varepsilon)\simeq \sqrt{27} r_H/2$, and only a small fraction of the particles fall into the black hole. However, in this case, the internal energy of each absorbed particle is high, and thus although a small fraction of the particles are absorbed at high temperatures, each of the absorbed particles carries much more energy than in the low temperature case, giving rise to a large energy accretion rate. The behavior of $\dot{\cal E}$ displayed in the right panel of  Fig.~\ref{Fig:AccretionRate} is the result of these two competing effects (the smaller fraction of absorbed particles versus higher internal energy for each gas particles as the temperature goes up).

\begin{figure}[h]
\begin{center}
\includegraphics[width=10cm]{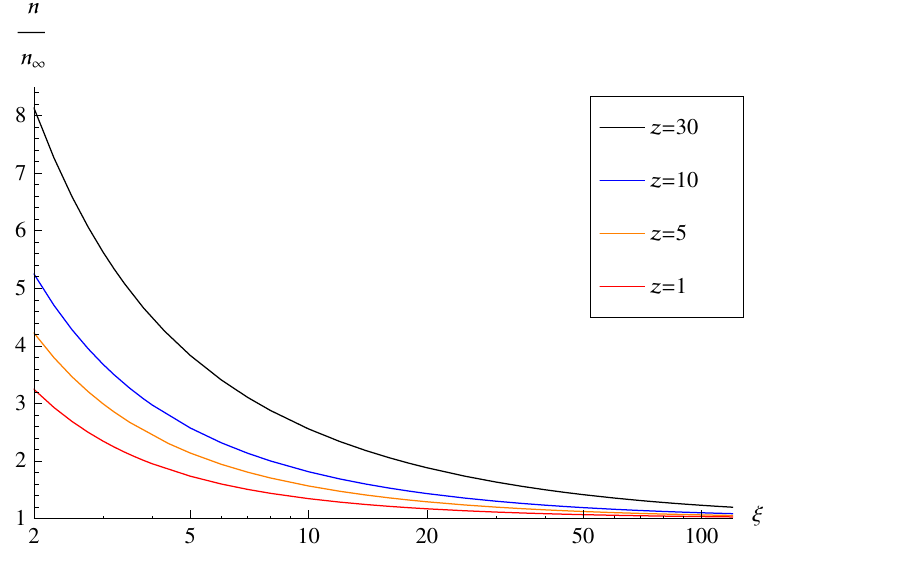}
\caption{\label{Fig:CompressionRate} The behavior of the compression rate as a function the dimensionless radial coordinate $\xi$ for different temperatures.}
\end{center}
\end{figure}
To reinforce this interpretation, we compute the compression rate $n/n_\infty$ and the energy density $\rho$ of the kinetic gas as a function of $\xi$ at different temperatures. In Fig.~\ref{Fig:CompressionRate}, we show the behavior of the compression rate $n/n_\infty$. Here, the particle density is defined invariantly by
\begin{equation}
n(\xi) = \sqrt{-J^\mu(\xi) J_\mu(\xi)},
\end{equation}
with $J^\mu$ the current density whose explicit representation can be found in Section IV.C in~\cite{pRoS16}. As can be observed from the plots in Fig.~\ref{Fig:CompressionRate}, at fixed temperature the compression rate increases as one moves closer to the black hole, as expected. However, at any fixed position, the compression rate decreases as the temperature increases. This further illustrates the fact that at high temperatures, a smaller fraction of the particles is absorbed by the black hole.

\begin{figure}[ht]
\centerline{\resizebox{9.0cm}{!}{\includegraphics{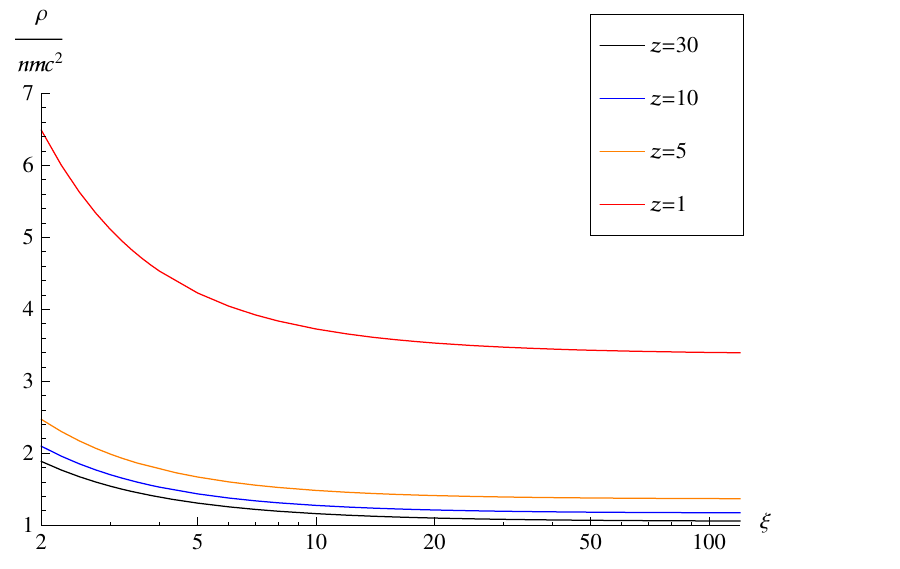}}
\hspace{-1.0cm}
\resizebox{9.0cm}{!}{\includegraphics{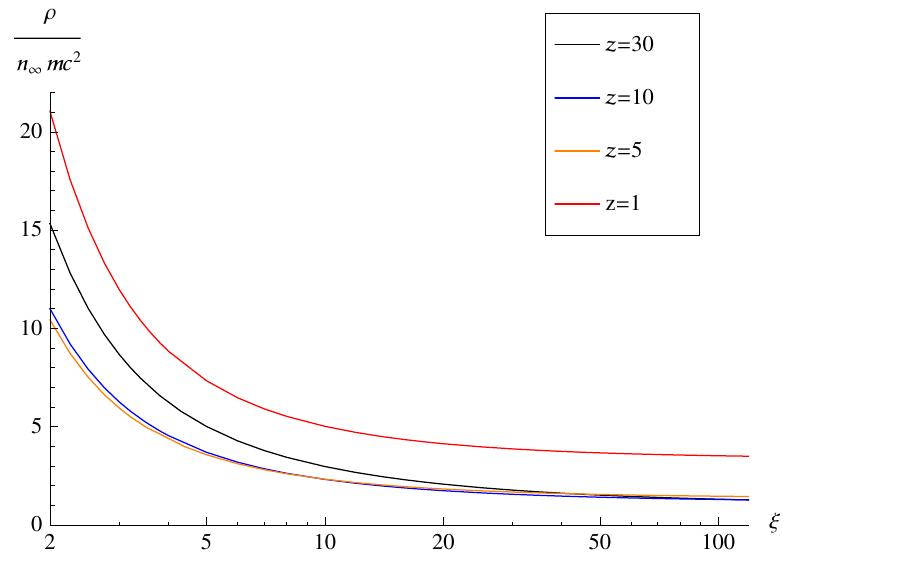}}}
\caption{\label{Fig:Energy} The behavior of the energy density $\rho$ as a function of $\xi$ at different temperatures. Left panel: Ratio between $\rho(\xi)$ and the rest energy density $n(\xi) m c^2$. Right panel: Ratio between $\rho(\xi)$ and the rest energy density $n_\infty m c^2$ at infinity.}
\end{figure}
Next, in Fig.~\ref{Fig:Energy} we show the behavior of the energy density $\rho(\xi)$ of the gas, compared to both its rest energy density at $\xi$ and the rest energy density at infinity. Here, $\rho(\xi)$ is defined invariantly as minus the eigenvalue corresponding to the unique future-directed, unit timelike eigenvector ${\bf e}_0$ of $T^\mu{}_\nu(\xi)$, such that $T^\mu{}_\nu e_0^\nu = -\rho e_0^\mu$. $\rho$ is computed numerically from the explicit expressions for $T^\mu{}_\nu(\xi)$ given in Section IV.C of Ref.~\cite{pRoS16}. We see from the plots in the left panel that as expected, the quantity $\rho(\xi)/(n(\xi) mc^2)$ (being equal to one plus the fraction between the internal energy of the gas and its rest energy) increases as one moves closer to the black hole or as the temperature goes up. In contrast to this, the behavior of the ratio between $\rho(\xi)$ and the rest energy density at infinity, $n_\infty m c^2$, is not monotonous in the temperature anymore, as is visible from the plots in the right panel of Fig.~\ref{Fig:Energy}. As explained before, the reason for this is the smaller fraction of particles being accreted as the temperature goes up, leading to the difference between the behavior of $n/n_\infty$ and $\rho/(n mc^2)$ (the first ratio decreases while the second one increases as temperature increases).

\begin{figure}[ht]
\centerline{\resizebox{8.5cm}{!}{\includegraphics{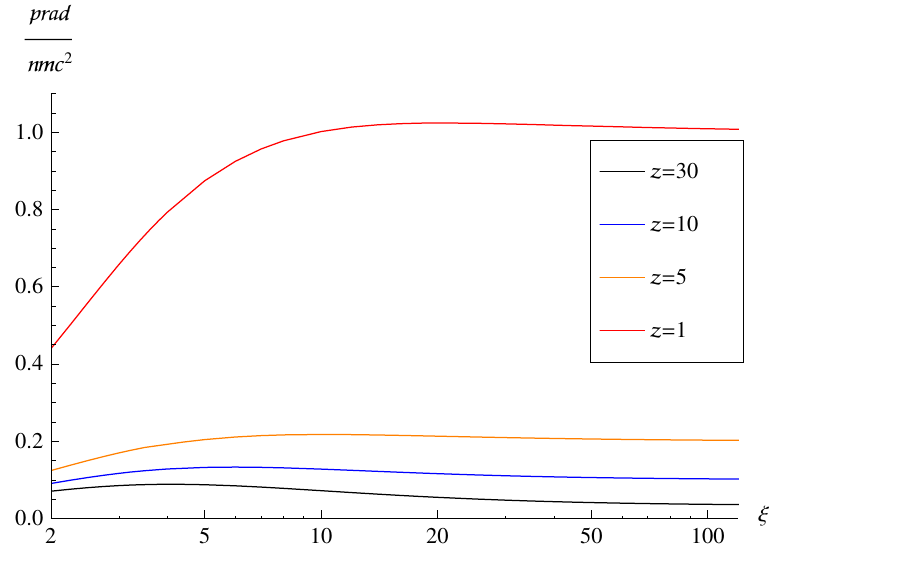}}\resizebox{8.5cm}{!}{\includegraphics{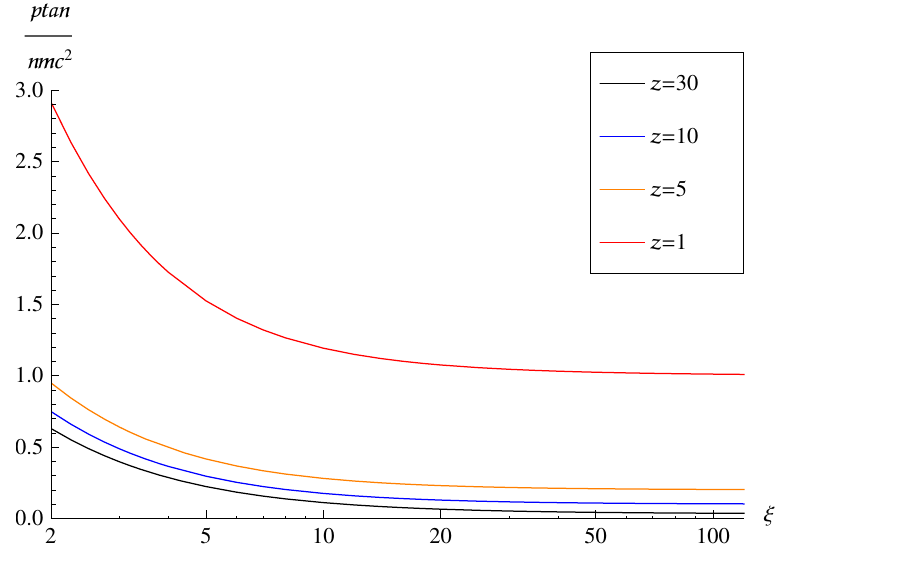}}}
\caption{The ratio between the radial (left panel) and tangential (right panel) pressures and the rest energy density as a function of $\xi$ for different values of the temperature.}
\label{Fig:PtanPrad}
\end{figure}

\begin{figure}[h]
\begin{center}
\includegraphics[width=10cm]{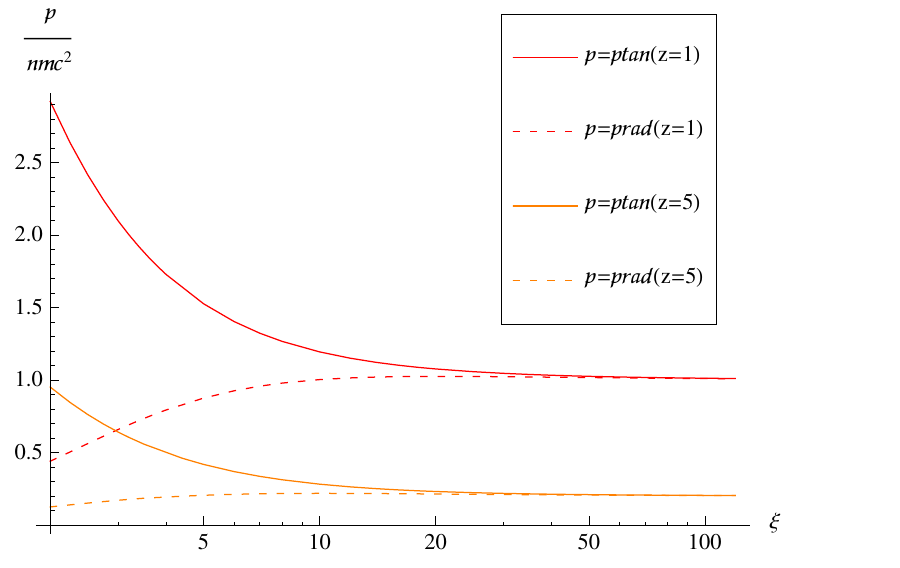}
\caption{\label{Fig:Comparison} A comparison between the radial and tangential pressure profiles for two different values of $z$. The solid lines refer to the tangential pressure, the dashed lines to the radial one. While the ratio $p_{tan}/p_{rad}$ converges to one as $\xi\to \infty$, it increases as one approaches the horizon, reaching a value of about $5$ for $z=1$ and about $8$ for $z=5$.}
\end{center}
\end{figure}

Next, we compute the radial and tangential pressures $p_{rad}$ and $p_{tan}$ of the gas. As already mentioned in the introduction, it was shown in~\cite{pRoS16} that at the horizon and in the low temperature limit $z\to \infty$ the relativistic kinetic gas behaves very differently than an isotropic perfect fluid, $p_{tan}$ being almost an order of magnitude larger than $p_{rad}$. In the following, we analyze the behavior of the pressures as a function of $\xi$ for finite temperatures. The radial and tangential pressures are obtained from the eigenvalues of $T^\mu{}_\nu$ corresponding to the spacelike eigenvectors; more specifically,
$$
T^\mu{}_\nu e_r^\nu = p_{rad} e_r^\mu,\qquad
T^\mu{}_\nu e_\vartheta^\nu = p_{tan} e_\vartheta^\mu,
$$
with ${\bf e}_r$ the unit outgoing radial eigenvector of $T^\mu{}_\nu$ and ${\bf e}_\vartheta := r^{-1}\partial_\vartheta$. In Fig.~\ref{Fig:PtanPrad} we show the behavior of these quantities divided by the rest mass energy $n mc^2$. For large values of $\xi$, the ratios $p_{rad}/(n mc^2)$ and $p_{tan}/(n mc^2)$ are approximately equal to each other, and they only depend on the temperature, consistent with an isotropic gas satisfying the ideal gas equation of state. However, as one approaches the horizon, $p_{tan}/(n mc^2)$ increases steadily while $p_{rad}/(mc^2)$ decreases close to the horizon, resulting in an anisotropic distribution of the pressure and a departure from the isotropic perfect fluid case. The ratio between $p_{tan}$ and $p_{rad}$ at the horizon varies between a factor of about $5$ for $z=1$ and about $10$ for large $z$, see Fig.~\ref{Fig:Comparison} and Ref.~\cite{pRoS16} for the case $z\to\infty$.

\begin{figure}[h]
\begin{center}
\includegraphics[width=10cm]{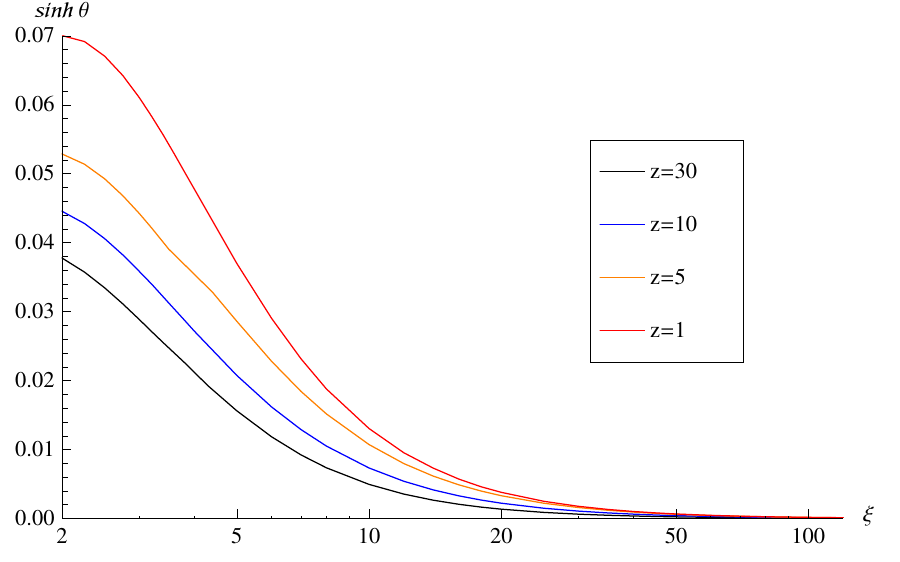}
\caption{\label{Fig:Angle} The quantity $\sinh\theta$ as a function of $\xi$ for different values of the temperature. While $\sinh\theta\to 0$ as $\xi\to \infty$ for all values of $z$, $\sinh\theta$ is clearly different from zero close to the horizon, implying that the kinetic gas does not behave as a perfect fluid in the vicinity of the horizon.
}
\end{center}
\end{figure}

As discussed in~\cite{pRoS16}, apart from the difference between the radial and tangential pressures, there is another property of the kinetic gas that distinguishes it from an isotropic perfect fluid configuration. In the latter, the current density is proportional to the four-velocity of the fluid which is also a timelike eigenvector of $T^{\mu}{}_\nu$. However, in the former case, ${\bf J} = J^\mu\partial_\mu$ does not coincide with the timelike eigenvector of $T^\mu{}_\nu$ in general. To quantify this difference, we introduce the hyperbolic angle $\theta$ defined by
$$
\cosh\theta = -g({\bf e}_0,{\bf u}),\qquad
\sinh\theta = g({\bf e}_r,{\bf u}),
$$
where ${\bf u} := {\bf J}/n$ is the mean four-velocity of the particles. In the perfect fluid case, ${\bf u}$ coincides with ${\bf e}_0$ and $\theta = 0$; hence the angle $\theta$ measures the deviation from the perfect fluid case. In Fig.~\ref{Fig:Angle} we plot $\sinh\theta$ as a function of $\xi$ for different temperatures. While asymptotically for $\xi\to \infty$ this quantity converges to zero for all values of $z$, it increases steadily as one approaches the horizon. In the limit $z\to\infty$, the value of $\sinh\theta$ at the horizon can be computed from the results in Section V.A of Ref.~\cite{pRoS16}, giving $\sinh\theta\simeq 0.0345$. As the temperature increases, this value increases until it reaches $\sinh\theta\simeq 0.07$ when $z=1$.

\section{Comparison with the isotropic perfect fluid case}

In this section, we wish to address the question of why our relativistic kinetic gas model behaves so much differently than the hydrodynamic model of Bondi-Michel accretion, even though in both cases the gas behaves exactly as an isotropic perfect fluid at infinity. Naively, one might have expected that in some limit the kinetic model should have reproduced the hydrodynamic flow, since the hydrodynamic equations can be obtained as an appropriate limit of a kinetic gas which is in (local) thermodynamic equilibrium~\cite{Huang-Book,wI63,Stewart-Book}. Although in our model we have neglected collisions, one could still argue that our distribution function in Eq.~(\ref{Eq:SphEquilSol}) corresponds to an equilibrium distribution function and hence, the corresponding observables should be fluid-like. So why is the accretion flow in our model different than in the hydrodynamic case?

\begin{figure}[h]
\begin{center}
\includegraphics[width=9cm]{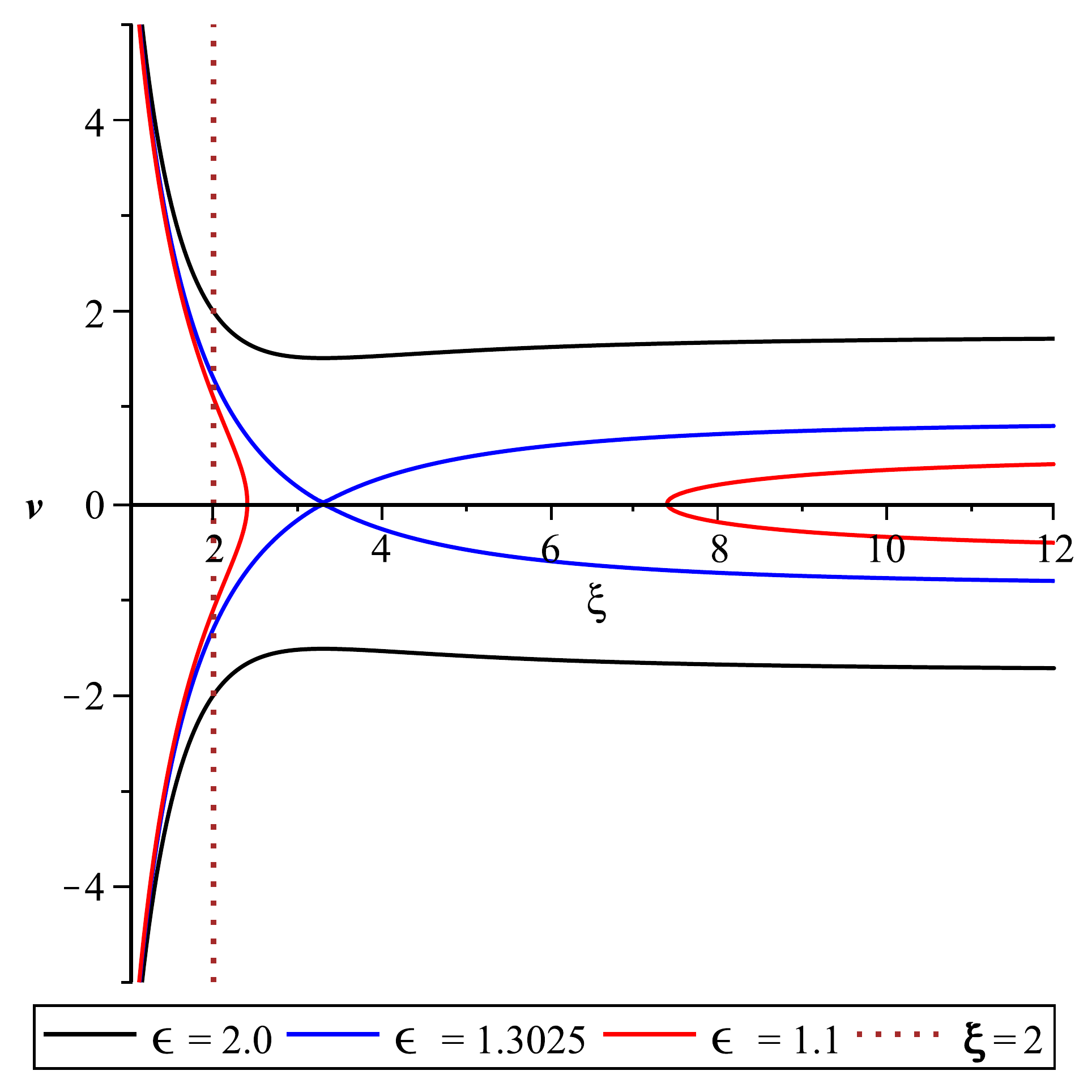}
\caption{\label{Fig:trajectories} (adapted from Fig.~2 in~\cite{pRoS16}) Projection of the phase diagram onto to $(\xi,v=\dot{r}/mc)$-plane illustrating the particle trajectories for different energy levels $\varepsilon$ and total dimensionless angular momentum $\lambda = 6$. The black curve in the region $v < 0$ corresponds to an incoming particle from infinity which is absorbed by the black hole, while the black curve in the region $v > 0$ describes an outgoing particle that is emitted from the white hole and escapes to infinity. The red curve in the region $\xi > 6$ describes a particle that is incoming from infinity but has large enough angular momentum $\lambda > \lambda_c(\varepsilon)$ to be reflected at the potential barrier, and the red curve in the region $\xi < 3$ describes a particle that is emitted by the white hole, is reflected at the potential barrier and absorbed by the black hole. The blue curve describes the separatrix and corresponds to the energy level $\varepsilon$ such that $\lambda_c(\varepsilon) = \lambda$.
}
\end{center}
\end{figure}

The answer comes from the observation that although, formally, the distribution function described by Eq.~(\ref{Eq:SphEquilSol}) looks like an equilibrium distribution function, in fact it is not. The reason for this can be traced back to the curved geometry of the Schwarzschild background and its causal properties, and relies on the fact that in our model in~\cite{pRoS16} the support of the distribution function is restricted to the region $\Gamma_{accr}$ of phase space corresponding to particle trajectories originating from the reservoir. On the one hand, this means that only particles with large enough energies $E > mc^2$ are admitted in the gas. Although this eliminates the possibility of occupying bounded trajectories, it should be noted that the contributions from such trajectories would not affect the observables in the vicinity of the horizon, since they are confined to the region $r > 3r_H/2$ outside the photon sphere. On the other hand, the restriction to $\Gamma_{accr}$ also eliminates the possibility of occupying those particle trajectories that emanate from the white hole (see Fig.~\ref{Fig:trajectories}) and whose contributions would affect the observables close to the horizon. Therefore, since it is truncated to the subset $\Gamma_{accr}$, the distribution function~(\ref{Eq:SphEquilSol}) is {\em not} an equilibrium distribution function.

In the collisionless case, the restriction to $\Gamma_{accr}$ is appropriate, since if a particle trajectory is occupied at one instant, it is occupied for all times, such that eliminating the particle trajectories emanating from the white hole makes sense from a physical point of view. In fact, if one considers the distribution function~(\ref{Eq:SphEquilSol}) on the full phase space instead of restricting it to $\Gamma_{accr}$, one obtains an isotropic perfect fluid configuration with four-velocity $u$ parallel to the Killing vector field $k = \partial_t$. However, this configuration describes a static flow which fails to be regular at the horizon, and thus it does not describe a Bondi-Michel-like flow or any other physically well-defined solution.

Clearly, the situation changes when collisions are taken into account, since in this case  a binary collision may cause an unoccupied trajectory to become occupied after the collision takes place. Hence, in this case, it does not make sense to restrict the phase space to $\Gamma_{accr}$, and the bounded trajectories and those emanating from the white hole have to be taken into account. We intend to analyze the effects of collisions in future work.

\section{Conclusions}

In this article, extending previous work~\cite{pRoS16}, we have presented a detailed analysis of the properties of the spherical, steady-state accretion of a relativistic, collisionless kinetic gas into a Schwarzschild black hole. Assuming that in the asymptotic region the gas is isotropic and described by an equilibrium distribution function, we have computed the relevant physical quantities such as the accretion rate, particle density, energy density and radial and tangential pressures, and we have analyzed their behavior as a function of the temperature of the gas at infinity and/or the radial coordinate. In particular, we have shown that for given mass of the black hole and given particle density number at infinity, the energy accretion rate diverges when $z\to 0$ or $z\to \infty$, having a minimum when the ratio $z = mc^2/(k_B T)$ between the rest mass of the particles and their thermal energy is of the order one. We attributed this behavior to the competition between two different effects: the smaller fraction of particles being accreted by the black hole versus their higher internal energy as the temperature increases.

Further, we have analyzed the difference between the properties of the collisionless flow to those of an isotropic perfect fluid, as is assumed in the Bondi-Michel model. This difference manifests itself in at least three different ways. First, in the low temperature limit, the accretion rate in the collisionless case is much lower than in the fluid case, leading to mass accretion rates of the order of $10^{-21} M_\odot/yr$ for accretion of the ionized component of the interstellar medium by a stellar mass black hole of mass $10M_\odot$ under typical conditions~\cite{Shapiro-Book}. Second, in the collisionless case, although equal at infinity, the radial and tangential pressures differ from each other at finite radius (the difference being largest at the horizon) implying that the gas is anisotropic. Third, in the collisionless case, the current density does not necessarily agree with the timelike eigenvector of the stress energy-momentum tensor. This difference can be quantified by an hyperbolic angle $\theta$ whose behavior we have analyzed. This behavior shows that in the vicinity of the horizon, the collisionless kinetic gas is different from a perfect fluid.

Finally, we have provided an explanation for the fact that in our model the kinetic gas does not behave as an isotropic perfect fluid, although it does so at infinity. As we show, the origin for this effect is geometrical, the non-trivial causal structure of the Schwarzschild black hole leading to a discrimination between the particle trajectories that end at the black hole and those that emanate from the white hole. It should be interesting to investigate the effects introduced by collisions, and to check whether such effects diminish the differences reported in this article between the kinetic and isotropic perfect fluid models.

\section*{Acknowledgement}

We thank Dar\'io N\'u\~nez and Thomas Zannias for fruitful discussions and questions that motivated some of the discussions presented in this work. This investigation was partially supported by CONACyT Grants No. 577742 and No. 271904, and by a CIC Grant to Universidad Michoacana.

\section*{References}
\bibliographystyle{iopart-num}
\bibliography{../References/refs_kinetic}

\end{document}